# Discovery of the progenitor of the type Ia supernova 2007on

Rasmus Voss[1,2] & Gijs Nelemans[3]

[1]Max Planck Institute for Extraterrestrial Physics, Giessenbachstraße, 85748, Garching, Germany. [2]Excellence Cluster Universe, Boltzmannstrasse 2, 85748, Garching, Germany. [3]Department of Astrophysics, IMAPP, Radboud University, Toernooiveld 1, 6525 ED, Nijmegen, The Netherlands.

**Type Ia supernovae are exploding stars that are used to measure the accelerated expansion of the Universe[1,2] and are responsible for most of the iron ever produced[3]. Although there is general agreement that the exploding star is a white dwarf in a binary system, the exact configuration and trigger of the explosion is unclear[4], which could hamper their use for precision Cosmology. Two families of progenitor models have been proposed. In the first, a white dwarf accretes material from a companion until it exceeds the Chandrasekhar mass, collapses and explodes[5,6]. Alternatively, two white dwarfs merge, again causing catastrophic collapse and an explosion[7,8]. Hitherto it has been impossible to determine if either (or both) models are correct. Here we report the discovery of an object at the position of the recent type Ia supernova (2007on) in the elliptical galaxy NGC 1404 on pre-**

**supernova archival X-ray images. Deep optical images show no sign of this object. From this we conclude that the X-ray source is the progenitor of the supernova, which favours the accretion model for this supernova, although the host galaxy is older (6-9 Gyr) than the age at which the explosions are predicted in the accreting models.**

The two proposed progenitor models of type Ia supernovae are radically different. In the accreting model a prolonged phase of mass transfer precedes the explosion, often identified with the bright so-called supersoft X-ray sources, binary stars in which the mass transfer is believed to be just fast enough to sustain steady nuclear burning on the surface of the white dwarf[9]. In the merger model the mass growth is extremely rapid during the merger, but there is no mass transfer before the explosion, and it is expected that such progenitors will be extremely faint. However, it has been suggested that the mass growth is slowed down significantly by the rapid rotation of the system[10], easing the problems with the triggering of the explosions when the mass growth is too rapid[11]. In that case there would be thousands of years between the merger and the actual explosion and the progenitors might also be X-ray sources.

Attempts to distinguish between the models have been based on indirect methods. Calculations of the rate at which supernovae occur as a function of the age of the stellar population show that the mergers occur in populations of all ages while the accreting models tend to occur mostly at intermediate ages (~200 Myr to ~2 Gyr)[12,13], and from the statistics of observed supernovae there is a growing evidence for a two component model for the rates (one component proportional to star formation and one proportional to mass)[14,15]. Other attempts to constrain the progenitors have come from very detailed studies of the spectra of the supernovae[16,17], and attempts have been made to search for the companion stars in supernova remnants[18].  We have taken a different approach by searching for ways to directly detect the progenitor of a type Ia supernova in pre-supernova images of the position in the sky where the supernova occurred.

On November 5, 2007  supernova SN2007on was found in the outskirts of the elliptical galaxy NGC 1404[19]. Optical spectra of the supernova[20] showed that the supernova was of type Ia. The position of the supernova of R.A. =03h38m50s.9, Decl = -35°34'30" (J2000) is about 70" from the core of the host, corresponding to 8 kpc for a distance of 20 Mpc to NGC 1404[21].  SWIFT observations on November 11 detected the supernova in the optical/uv monitor but not in the X-ray telescope[22.] We analysed the SWIFT data and determined the position of the supernova as  R.A. = 03h38m50s.98, Decl. = -35°34'31".0 (J2000), with uncertainty of 1" (Fig. 1). The highest magnitude is V ≈ 13 (on November 16), yielding an absolute magnitude of Mv ≈ -18.5 making it a rather faint type Ia supernova[23]. NGC 1404 is an elliptical galaxy in the Fornax cluster and its colours suggest its stellar population is old, with estimates ranging from 6 to 9 Gyr (error about 2 Gyr) and no sign of recent star formation.[24,25]

We investigated the supernova position, using archival data prior to the explosion from the Hubble Space Telescope and Chandra X-ray Observatory, to search for a possible progenitor to the SNIa (Fig. 1).  In the *Chandra* observations obtained in 2003 a source is detected close to the position of the supernova, at the coordinates R.A. =03h38m50s. 91, Decl. = -35°34'30".9 (with a 1-sigma statistical error of 0".25, as well as a ~0.5" error on the absolute astrometric precision, based on correlations between  X-ray sources and the 2MASS and DSS images of the region). No optical source was detected to an absolute magnitude limit of -4.5. The X-ray source is 0.9(±1.3)" from the supernova, consistent with being its progenitor. The X-ray source is detected with 14.1(±4.6) counts, a 4.0 sigma significance using circular aperture photometry, and 5.0 sigma based on a wavelet analysis using the program wavdetect. Within a radius of 1' from the supernova position there are 7 detected sources, giving a density of detected sources of 2.2±0.8 per square arcminute. The probability of a chance coincidence of a source within a distance of 1.3" of the supernova position is therefore 0.3%. Even given the fact that this is not the first trial (the 4[th], see below), the likelihood of a chance alignment is very small. As globular clusters are known to be abundant in low mass X-ray binaries, the alignment of an X-ray source and the supernova would not be so significant if the supernova went off in a globular cluster. However this possibility is excluded by the non-detection of a source in the optical images. We therefore conclude that we have detected the progenitor of 2007on.

With the low number of counts it is impossible to determine the shape of the X-ray spectrum of the progenitor. Instead we investigate the source properties using the number counts in the source and background regions in three different energy bands, S (0.3-1.0 keV), M (1.0-2.0 keV) and H (2.0-8.0 keV). The results are S=12, M=4, H=5, with background expectations of 3.8±2.1, 2.0±1.4 and 1.1±1.2, respectively.  Analysis of all (8) individual images available in the *Chandra* archive shows no sign of variability. The luminosity of the source, for a distance of 20 Mpc is $3.3 \pm 1.5 \ 10^{37}$ erg/s (for a flat photon spectrum in each band, a power law with index 2 gives $2.2 \pm 0.9 \ 10^{37}$ erg/s).

In the archives there are 3 more SNe type Ia in galaxies with a distance of <25Mpc (all lenticular galaxies) and with Chandra observations longer than 10 ks prior to the SN explosion. We do not detect an X-ray source at the position of the supernova in any of them. The 3-sigma upper limits, using the same aperture method as above, and assuming the same spectral model are presented in Table 1, together with the parameters of 2007on.

The discovery of a luminous X-ray source that is likely to be the direct progenitor of 2007on has important consequences for our understanding of type Ia supernovae. The X-ray luminosity is fully consistent with the typical luminosities of supersoft sources[26], that are similar to the expectations from the accreting progenitor model. Also, their absolute magnitudes[26] are around -1 to -2, consistent with the non-detection in the optical images. If in the merger model the explosion immediately follows the actual merger the progenitor is not expected to emit X-rays before the supernova explosion, and this scenario is therefore inconsistent with the observed progenitor of SN2007on. However, if the lower-mass white dwarf is disrupted at the onset of Roche-lobe overflow, and forms a long-lived disk around the more massive white dwarf[10] the merged object may be a strong X-ray source before the explosion. Recent detailed calculations of these objects suggest that they would have X-ray luminosities about an order of magnitude lower than the progenitor we discovered[27]. We therefore conclude that our result favours the accreting model. Alternatively, very high accretion rates in the early phases of the evolution of AM CVn systems can also lead to steady burning on white dwarfs, but now of accreted helium[28]. Although the rate of Ia supernovae from this channel is very low, it is consistent with the properties of the progenitor.

The spectrum of the progenitor is relatively hard, compared to the typical supersoft sources that have temperatures below 100 eV. A 100 eV black body model can be ruled out, but models with 200-300 eV and a modest intrinsic absorption of $10^{21}$ atoms/cm$^2$ are consistent with the observed counts. In addition the age of the stellar population of NGC 1404 could pose a problem for the supersoft source interpretation: the models predict life times only up to about 2 Gyr[11,12], quite a bit younger than the inferred age of the population of NGC 1404. Maybe this indicates the progenitor was a lower-mass system, such as a symbiotic binary, for which recently hard X-rays have been discoved[29]. However, it can not be excluded that a small population of younger stars is present in the host galaxy of 2007on.

Our discovery opens a new method for the study of type Ia supernova progenitors. This first detection favours the accreting model and the three upper limits on previous supernovae are not strong enough to put additional constraints. However, this does not prove that the other models cannot lead to type Ia supernovae. Searches for supersoft sources in nearby galaxies have resulted in many fewer supersoft sources than expected and needed to explain all type Ia supernovae[30], even though the strong variability of these sources complicates the analysis. Also the growing support for a two component model for the rate of supernovae may very well indicate the complementarity of both progenitor models. Future detections or strong upper limits on pre-supernova X-ray luminosities are important for the understanding of this issue.

*Acknowledgements:* We thank the Central Bureau for Astronomical Telegrams for providing a list of supernovae. This research has made use of data obtained from the Chandra Data Archive and software provided by the Chandra X-ray Center in the application package CIAO, and of Swift data obtained from the High Energy Astrophysics Science Archive Research Center (HEASARC), provided by NASA's Goddard Space Flight Center. The observations from the NASA/ESA Hubble Space Telescope were obtained from the data archive at the Space Telescope Institute. We thank NOVA for support. G.N. is supported by an NWO VENI grant.

*Author contributions:* The authors have contributed equally to the paper

**Table 1 Nearby SNeIa observed with Chandra before the explosion**

| Supernova | 2007on | 2006mr | 2004W | 2002cv |
|---|---|---|---|---|
| Galaxy | NGC 1404 | NGC 1316 | NGC 4649 | NGC 3190 |
| Galaxy Type | Elliptical | Lenticular (Sa) | Lenticular (S0) | Lenticular (Sa) |
| Distance | 20 Mpc | 18.1 Mpc | 15.9 Mpc | 22.4 Mpc |
| OBS-ID | 2942 & 4174 | 2022 | 785 | 2760 |
| Time before SN | ~4 years | ~5 months | ~4 years | ~2 months |
| Count rate | $1.9\ (\pm0.6)\ 10^{-4}$ | $<9.2\ 10^{-4}$ | $<3.2\ 10^{-4}$ | $<3.6\ 10^{-4}$ |
| Luminosity | $3.3\ (\pm1.5)\ 10^{37}$ erg/s | $<1.3\ 10^{38}$ erg/s | $<3.5\ 10^{37}$ erg/s | $<7.9\ 10^{37}$ erg/s |

Fig. 1

Images of the region around SN2007on.

The images have been taken by the Chandra X-ray observatory (top), Hubble Space Telescope (middle) and Swift (bottom), and are smoothed by a gaussian with a 2 pixel FWHM. The circle gives a 5" radius around the position of the supernova (the circle is much larger than the positional uncertainty). The central parts of NGC 1404 are visible south of the supernova position.

There are 8 archival Chandra imaging observations with the supernova position within the field of view.  Only 2 of the observations (OBS-ID 2942 and 4174 taken in 2003 on 02/13 and 05/28) with a combined exposure time of 74.9 ks have telescope pointings within 6' from the SN position and thus enough sensitivity to be useful. We added the two observations, and the resulting image is shown in the top panel. Within an aperture of 2" the X-ray source has 21 counts, whereas a background region in the annulus 2"-10" has 165 counts giving an expectation value of 6.9 counts in the source region. The poisson probability of having 21 or more counts with this background level is $1.12 \ 10^{-5}$ and the source is therefore detected with a significance of 4.0 and the number of source counts is 14.1 (±4.6).

In the HST archive there are four WFPC2, four ACS and 1 NICMOS images covering the position of the supernova. No source is present within 2" of the given position in any of the images. The deepest (ACS) images in the F475W (760 sec, taken 2004-09-10) and F814W (1224 sec, taken 2006-08-06)  bands have a limiting magnitude of about 27 (Fig. 1), corresponding to an absolute magnitude of -4.5. The image with the F475W filter is shown in the middle panel. In the bottom panel a Swift UVOT V-band image (taken Nov. 11, 2007) of the supernova is shown.

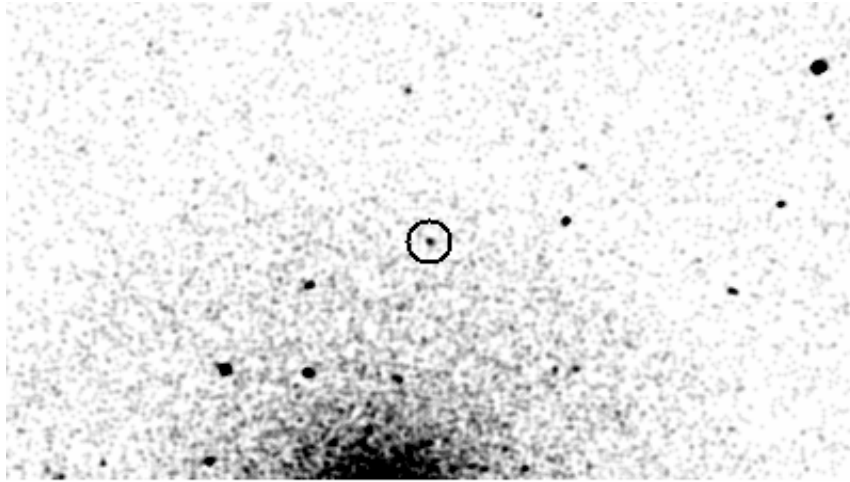

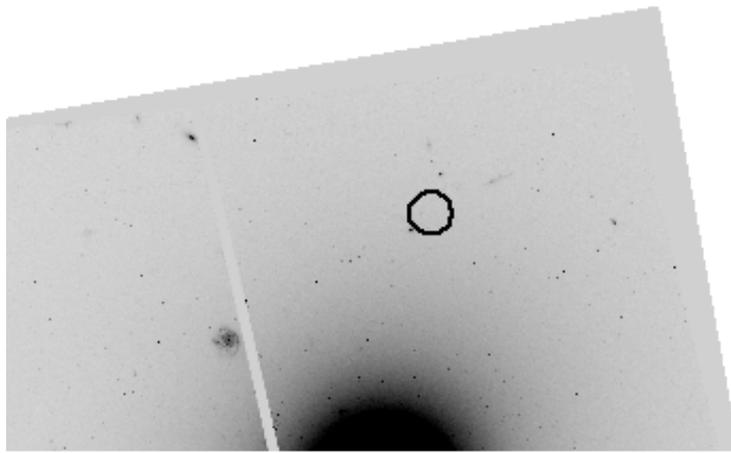

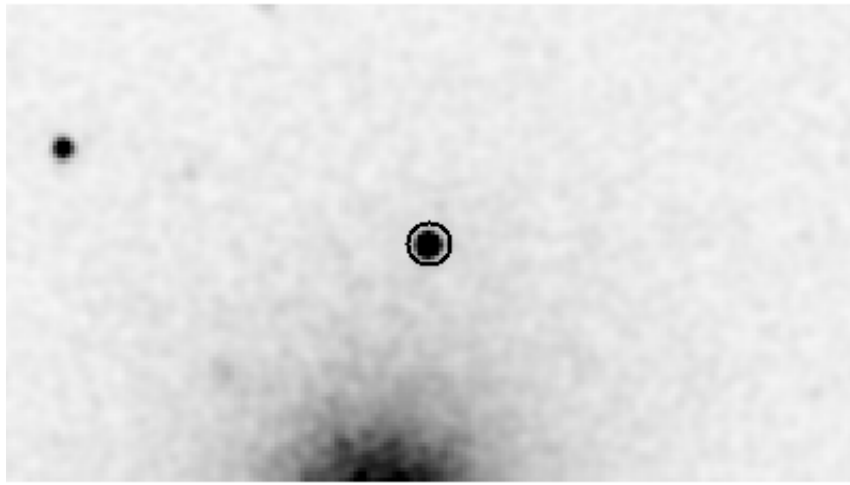